# Two-dimensional electron gas in a modulation-doped SrTiO$_3$/Sr(Ti,Zr)O$_3$ heterostructure

**Adam P. Kajdos[1], Daniel G. Ouellette[2], Tyler A. Cain[1], and Susanne Stemmer[1]**

[1]Materials Department, University of California, Santa Barbara, California, 93106-5050, USA.

[2]Department of Physics, University of California, Santa Barbara, California, 93106-9530, USA.

**Abstract**

A two-dimensional electron gas (2DEG) in $SrTiO_3$ is created via modulation doping by interfacing undoped $SrTiO_3$ with a wider-band-gap material, $SrTi_{1-x}Zr_xO_3$, that is doped n-type with La. All layers are grown using hybrid molecular beam epitaxy. Using magnetoresistance measurements, we show that electrons are transferred into the $SrTiO_3$, and a 2DEG is formed. In particular, Shubnikov-de Haas oscillations are shown to depend only on the perpendicular magnetic field. Experimental Shubnikov-de Haas oscillations are compared with calculations that assume multiple occupied subbands.

Two-dimensional electron gases (2DEGs) at interfaces between SrTiO$_3$ and other complex oxides exhibit unique phenomena, including superconductivity [1], Rashba spin-orbit coupling [2], and magnetism [3-7]. Most studies thus far have focused on 2DEGs at polar/nonpolar interfaces, such as LaAlO$_3$/SrTiO$_3$ [8-10], LaTiO$_3$/SrTiO$_3$ [11,12] and GdTiO$_3$/SrTiO$_3$ [13,14]. Mobile carriers in these heterostructures arise from a polar discontinuity at the interface, which gives rise to extremely high carrier densities on the order of ~$3\times10^{14}$ cm$^{-2}$ (LaAlO$_3$/SrTiO$_3$ interfaces typically have lower densities, for reasons that are not yet well understood [9]). An alternative route to high-mobility 2DEGs is modulation doping, originally developed for III-V heterostructures [15], leading to devices such as high-electron mobility transistors [16,17], and scientific discoveries, such as the fractional quantum Hall effect [18]. This approach spatially separates the mobile charge from the ionized dopants by transferring them into an undoped layer. A heterostructure that has suitable conduction band alignments is therefore essential for modulation doping.

Here we show that a 2DEG in SrTiO$_3$ can be created by modulation doping. We use SrTi$_{1-x}$Zr$_x$O$_3$ as the doped, wider band gap perovskite, i.e., the analog of AlGaAs in GaAs/AlGaAs structures. SrTi$_{1-x}$Zr$_x$O$_3$ is a solid solution of cubic SrTiO$_3$ and orthorhombic SrZrO$_3$. The band gap of SrZrO$_3$ (5.6 eV [19]) is substantially larger than that of SrTiO$_3$ (3.2 eV). The band alignment between SrTiO$_3$ and SrZrO$_3$ is Type I, with a conduction band offset, $\Delta E_c$, of 1.9 eV [20]. Given the large $\Delta E_c$, reasonable band offsets should be achievable even for small $x$, which serve to reduce the substantial lattice mismatch between SrTiO$_3$ ($a$ = 3.905 Å) and SrZrO$_3$ (pseudocubic unit cell, $a \approx 4.1$ Å), and facilitate doping, which would likely be difficult in a wide band gap material such as SrZrO$_3$.

Figure 1 shows a sketch of a modulation doped SrTiO$_3$/SrTi$_{0.95}$Zr$_{0.05}$O$_3$ heterostructure, along with a band diagram of the conduction band edge, Fermi level and the electron density, as calculated using a 1D Poisson solver [21]. A $\Delta E_c$ of 95 meV was estimated using Vegard's law. The SrTi$_{0.95}$Zr$_{0.05}$O$_3$ layer is n-type, doped with La to a carrier density of $2\times10^{19}$ cm$^{-3}$. A 2DEG resides in the SrTiO$_3$.

Experimentally realizing this structure requires the growth of epitaxial, stoichiometric SrTiO$_3$ and SrTi$_{1-x}$Zr$_x$O$_3$ thin films, to ensure high mobility, avoid charge trapping, and allow for doping of the SrTi$_{1-x}$Zr$_x$O$_3$. Here we use hybrid molecular beam epitaxy (MBE), previously developed for high-mobility SrTiO$_3$ [22,23]. In this approach, both Ti and Zr are supplied from high-purity (99.999%) metal-organic precursors, titanium tetra isopropoxide (TTIP) and zirconium tert-butoxide (ZTB), respectively.

The SrTi$_{0.95}$Zr$_{0.05}$O$_3$/SrTiO$_3$ heterostructure discussed in this paper consisted of a 120 nm undoped SrTiO$_3$ layer grown on (001) SrTiO$_3$, followed by growth of a 30-nm-thick La-doped SrTi$_{0.95}$Zr$_{0.05}$O$_3$ layer (see Fig. 1). For SrTi$_{1-x}$Zr$_x$O$_3$, both precursors were co-evaporated, and the beam equivalent pressures were adjusted relative to that of Sr to give A:B site stoichiometric films (where A = Sr and B = Ti or Zr) of the desired Zr content ($x$). Growth optimization to obtain the correct A:B site ratio closely followed our approach for SrTiO$_3$ [23], and details will be reported elsewhere. Oxygen was supplied by an rf plasma source. The substrate temperature was 900 °C as measured by thermocouple. La was evaporated from a solid source effusion cell to obtain a carrier concentration of $2\times10^{19}$ cm$^{-3}$. Doping calibrations were carried out using thick, uniformly doped layers, which confirmed *n*-type doping of the $SrTi_{0.95}Zr_{0.05}O_3$ layers, but showed much lower mobilities than the modulation doped samples investigated here. In-situ reflection high-energy diffraction (RHEED) patterns were streaky during and after growth for all films and showed periodic oscillations in intensity at the beginning of growth, indicating layer-by-layer growth [24]. 2θ-ω x-ray diffraction scans in the vicinity of the $SrTiO_3$ 002 reflection showed a single, separate 002 peak for the $SrTi_{1-x}Zr_xO_3$ film and clearly defined Laue oscillations [24]. The composition *x* was calibrated by comparing the unstrained unit cell volume, obtained from lattice parameter measurements, to that of bulk $SrTi_{1-x}Zr_xO_3$ [25]. The sample was post-growth annealed in a rapid thermal annealing furnace in 1 atm of oxygen at 800 °C for 30 s to ensure oxygen stoichiometry. Hall and longitudinal resistance measurements at temperatures between 300 K and 2 K and magnetic fields (*B*) up to 14 T were performed in van der Pauw geometry using a Physical Properties Measurement System (Quantum Design). Ohmic contacts (40nm-Al/20nm-Ni/400nm-Au) were deposited by electron beam evaporation through a shadow mask onto the corners of a square sample. The longitudinal magnetoresistance was measured using a standard lock-in technique in a $He^3$ cryostage at temperatures ranging from 0.45 K to 3 K. Data at low B showed positive magnetoresistance at low temperatures, with a sharp dip near *B* = 0, possibly due to weak antilocalization. Magnetoresistance measurements as a function of the angle *θ* that *B* makes with the surface normal were performed at 2 K using a horizontal sample rotator stage (*θ* = 0° indicates that *B* is perpendicular to the interface).

To confirm the existence of a 2DEG and to probe the dimensionality of the electron system in the $SrTi_{0.95}Zr_{0.05}O_3/SrTiO_3$ heterostructure, we investigate Shubnikov-de Haas oscillations that appear in the longitudinal magnetoresistance in a quantizing magnetic field. In particular, for a 2DEG, the periodicity of these oscillations, which are due to quantization into Landau levels, should only depend on the component of $B$ that is normal to the interface.

Figure 2 shows the longitudinal magnetoresistance at different temperatures. At 0.45 K, Shubnikov-de Haas oscillations are observed for $B > 5$ T. Oscillations persist to temperatures up to 3 K. The oscillating component of the magnetoresistance, $\Delta R_{xx}$, was obtained by subtracting the non-oscillating background using multiple polynomial fits. Figure 3(a) shows $\Delta R_{xx}$ as a function of $1/B$ [$\Delta R_{xx}(1/B)$] at 0.45 K. A Fourier transform (FT) of $\Delta R_{xx}(1/B)$ is shown in the inset and exhibits multiple peaks, which will be further discussed below. Figure 3(b) shows Shubnikov-de Haas oscillations for different $\theta$, measured at 2 K and plotted against $(B\cos\theta)^{-1}$. Resistance maxima and minima appear at the same values of $(B\cos\theta)^{-1}$ for $\theta$ up to 60°, confirming the two-dimensionality of the electron system. Therefore, as suggested by the band diagram in Fig. 1, a 2DEG has been created on the SrTiO$_3$-side of the interface.

Shubnikov-de Haas oscillations allow for extracting important characteristics, including the effective masses ($m^*$) for the subbands and the quantum scattering time ($\tau_q$) or Dingle temperature ($T_D$). For 2DEGs with a single subband, values for $m^*$ and $T_D$ are determined from the decay in the amplitude of oscillations with increasing temperature and by constructing a Dingle plot at a fixed temperature, respectively. Multiple subbands (or spin splitting) complicate such analyses, since oscillations contain contributions from several components. Therefore, we fit the experimental data, $\Delta R_{xx}(1/B)$, at 0.45 K to the standard equation [26]:

$$\frac{\Delta R_{xx}}{R_0} = \sum_{i=1}^{l} 2A_i \exp\left(-\frac{2\pi^2 k_B T_{D,i}}{\hbar \omega_c}\right) \frac{X}{\sinh(X)} \cos\left(\frac{2\pi f_i}{B} + \pi\right), \qquad (1)$$

where $i$ is the subband index, $l$ the number of occupied subbands, $A_i$ are amplitude factors associated with intrasubband scattering probabilities, $k_B$ is the Boltzmann constant, $T_{D,i}$ is the Dingle temperature of each subband, $\hbar$ is the reduced Planck's constant, $\omega_c$ is the cyclotron frequency, $f_i$ are the oscillation frequencies (in $1/B$), and $X = (2\pi^2 k_B T)/(\hbar \omega_c)$. Figure 4 shows experimental and calculated $\Delta R_{xx}(1/B)$ at temperatures between 0.45 K and 3 K for $l = 4$. The temperature-dependence was calculated and the $m^*$'s adjusted to match the experimentally observed behavior. The parameters obtained from the fits and the extracted sheet carrier density ($n_s$), $\tau_q$, and quantum mobility ($\mu_q$) for each subband are shown in Table I. The frequencies are in good agreement with those obtained in the FTs shown in Fig. 3(a). The lowest value for $m^*$ (0.95 $m_0$, where $m_0$ is the free electron mass) is consistent with a $d_{xy}$-derived subband [27-30]. Spin-orbit coupling can hybridize $d_{xy}$-derived with $d_{xz}/d_{yz}$-derived subbands [29]. This results in a heavier in-plane mass. Theoretical calculations support multiple occupied subbands at this carrier density [29]. Alternative models and fits to the experimental Shubnikov-de Haas oscillations were also explored (see [24]), including whether multiple frequencies could arise from two spin-split subbands. A good qualitative match with the data could be obtained, and at present we cannot distinguish between these models [24]. These uncertainties place limitations on the accuracy of the extracted values in Table I. The extracted $\mu_q$ of around 2000 cm$^2$V$^{-1}$s$^{-1}$ are consistent with the onset of magnetoresistance oscillations around 4-5 T, i.e., when $\mu_q B > 1$.

The total sheet density estimated from Table I is approximately $8.4 \times 10^{12}$ cm$^{-2}$, which is only 19% the Hall density, $n_{s,H} = 4.46 \times 10^{13}$ cm$^{-2}$, measured at 300 K. The remaining carriers can be reasonably assumed to reside in the La-doped SrTi$_{0.95}$Zr$_{0.05}$O$_3$ layer. Carrier distributions calculated in Fig. 1 confirm this picture qualitatively. Using reasonable assumptions for the mobilities and carrier densities in both layers, a Hall coefficient can be calculated from a multi-carrier model and compared with experimentally measured values. Good agreement is found [24]. The two spin-split subband interpretation of the data (see ref. [24]) results in lower carrier densities. Prior studies of 2DEGs in SrTiO$_3$ found upwards of 60-95% of carriers in the 2DEG to not to give rise to oscillations [14,31-34], which is likely due to disorder limiting the mobility of a significant fraction of the carriers, preventing oscillations to be resolved for all subbands.

Discrepancies exist between calculated and experimental $\Delta R_{xx}$ in all models. One possible explanation is intersubband scattering [35,36], which was neglected in Eq. (1). The energy spacing between subbands, $\Delta E$, is very small:

$$\Delta E = \frac{q \hbar \Delta f}{m^*}, \qquad (2)$$

where $\Delta f$ is the difference in oscillation frequencies between any two subbands, and $q$ the elementary charge. From Table I, $\Delta E$ ranges from 0.15 meV to 5 meV. Intersubband scattering may thus be significant. It is generally more significant at low $B$ [35,36] and could thus potentially affect finer features that would otherwise appear in the measured resistance. At present, a complete theory of magnetoresistance oscillations of 2DEGs in SrTiO$_3$ is still lacking.

In summary, we have shown that modulation doping provides an alternative route to 2DEGs in SrTiO$_3$. Future studies should be dedicated to a complete understanding of the quantum oscillations, and to further improving the mobility in these structures. For example, remote ionized impurity scattering can be mitigated by inserting an undoped SrTi$_{1-x}$Zr$_x$O$_3$ spacer. Lower-density 2DEGs can be obtained by optimizing the doping and Zr content of the SrTi$_{1-x}$Zr$_x$O$_3$, which determines the band offsets.

The authors thank David Awschalom for making available the equipment for magnetotransport measurements and Jim Allen, Guru Khalsa and Allan MacDonald for helpful discussions. A.P.K. and T.A.C. received support from the National Science Foundation through a Graduate Research Fellowship (Grant no. DGE-1144085) and the Department of Defense through a NDSEG fellowship, respectively. Experimental costs (A. P. K.) were supported by the UCSB MRL, which is supported by the MRSEC Program of the National Science Foundation under Award No. DMR-1121053 and (T. A. C.) by the Center for Energy Efficient Materials, an Energy Frontier Research Center funded by the DOE (Award No. DESC0001009). The work also made use of the UCSB Nanofabrication Facility, a part of the NSF-funded NNIN network.

**Table I:** Parameters extracted from fits of the experimental Shubnikov-de Haas oscillations measured at 0.45 K to a four-subband model [Eqn. (1)]. Also shown are the sheet carrier densities ($n_s$), quantum scattering times ($\tau_q$), and quantum mobilities ($\mu_q$) of each subband extracted from the fits. The effective mass $m^*$ is determined from the cyclotron frequency $\omega_c = eB/m^*$, $\tau_q$ is calculated from $\tau_q = \hbar/(2\pi k_B T_D)$, and $n_s$ is calculated from $f = nh/2q$, where $h$ is Planck's constant.

| Subband index $i$ | $R_0 A$ ($\Omega$) | $f$ (T) | $m^*$ ($m_0$) | $T_D$ (K) | $n_s$ (cm$^{-2}$) | $\tau_q$ (s) | $\mu_q$ (cm$^2$V$^{-1}$s$^{-1}$) |
|---|---|---|---|---|---|---|---|
| 1 | 0.8 | 68.4 | 1.2 | 0.8 | $3.30 \times 10^{12}$ | $1.52 \times 10^{-12}$ | 2225 |
| 2 | 6.5 | 43 | 1.5 | 1.1 | $2.08 \times 10^{12}$ | $1.10 \times 10^{-12}$ | 1288 |
| 3 | 2 | 41.2 | 1.4 | 0.8 | $1.99 \times 10^{12}$ | $1.52 \times 10^{-12}$ | 1907 |
| 4 | 2 | 21.2 | 0.95 | 0.9 | $1.02 \times 10^{12}$ | $1.35 \times 10^{-12}$ | 2496 |

**Figure captions**

**Figure 1:** Schematic of a modulation-doped SrTi$_{0.95}$Zr$_{0.05}$O$_3$/SrTiO$_3$ heterostructure and corresponding band diagram. The conduction band offset was estimated using experimental values for SrZrO$_3$/SrTiO$_3$ [20] and assuming Vegard's law. The conduction band edge ($E_c$), Fermi level ($E_f$), and carrier density (*n*) were calculated using a 1D Poisson solver [21]. The dopant density in the SrTi$_{0.95}$Zr$_{0.05}$O$_3$ and the surface pinning potential were assumed to be $2\times10^{19}$ cm$^{-3}$ and 0.1 eV, respectively. The SrTiO$_3$ layer is nominally undoped with ~$10^{16}$ cm$^{-3}$ residual donor-type defects (an upper limit established by doping studies for these high-quality MBE films).

**Figure 2:** Longitudinal magnetoresistance ($R_{xx}$) measured at temperatures between 0.45 K and 3 K.

**Figure 3:** (a) Shubnikov-de Haas oscillations, $\Delta R_{xx}(1/B)$, measured at 0.45 K and $\theta = 0°$. The inset shows a Fourier transform of the data. Approximate positions for four peaks are labeled. (b) Shubnikov-de Haas oscillations measured at 2 K at various tilt angles $\theta$ (the angle between *B* and the interface plane normal) and plotted as a function of the inverse perpendicular component of the magnetic field, $(B\cos\theta)^{-1}$.

**Figure 4:** (a) Temperature-dependent Shubnikov-de Haas oscillations, $\Delta R_{xx}(1/B)$, measured at $\theta = 0°$ and at temperatures between 0.45 K and 3 K. (b) Calculated [Eq. (1)]

Shubnikov-de Haas oscillations for temperatures between 0.45 K to 3 K, using the parameters listed in Table I.

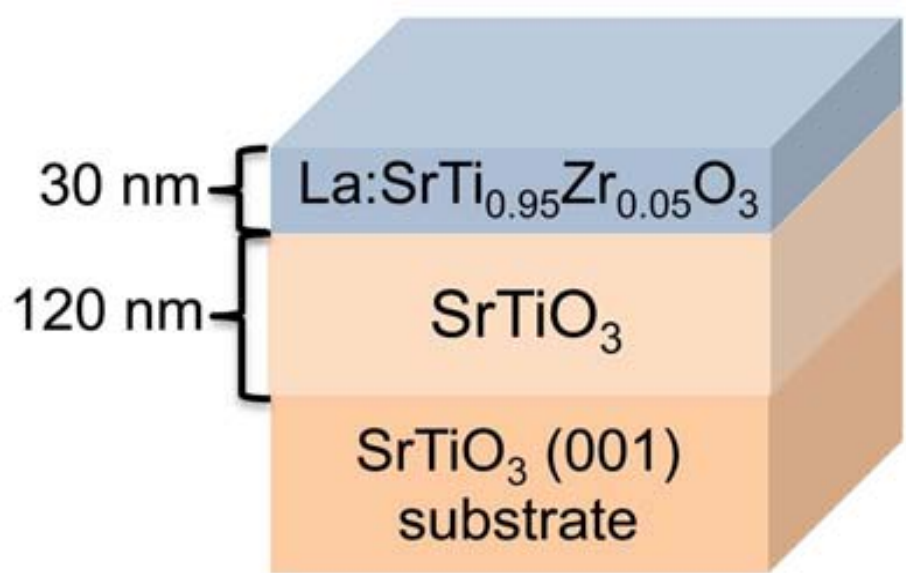
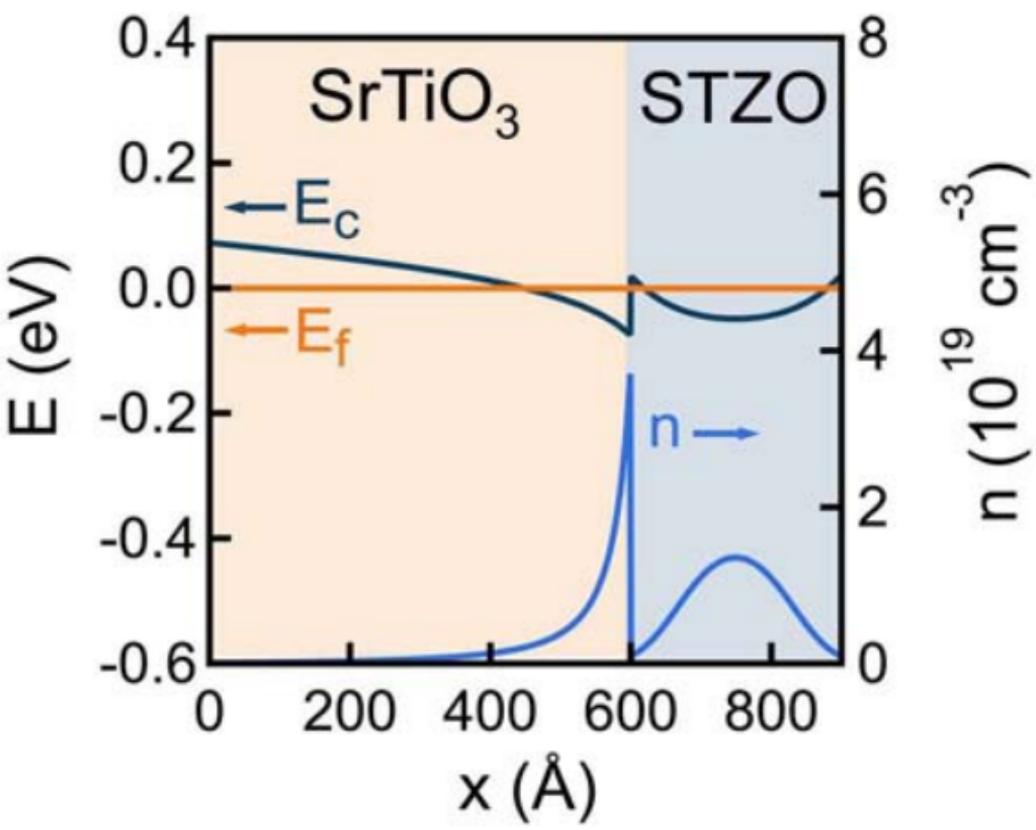

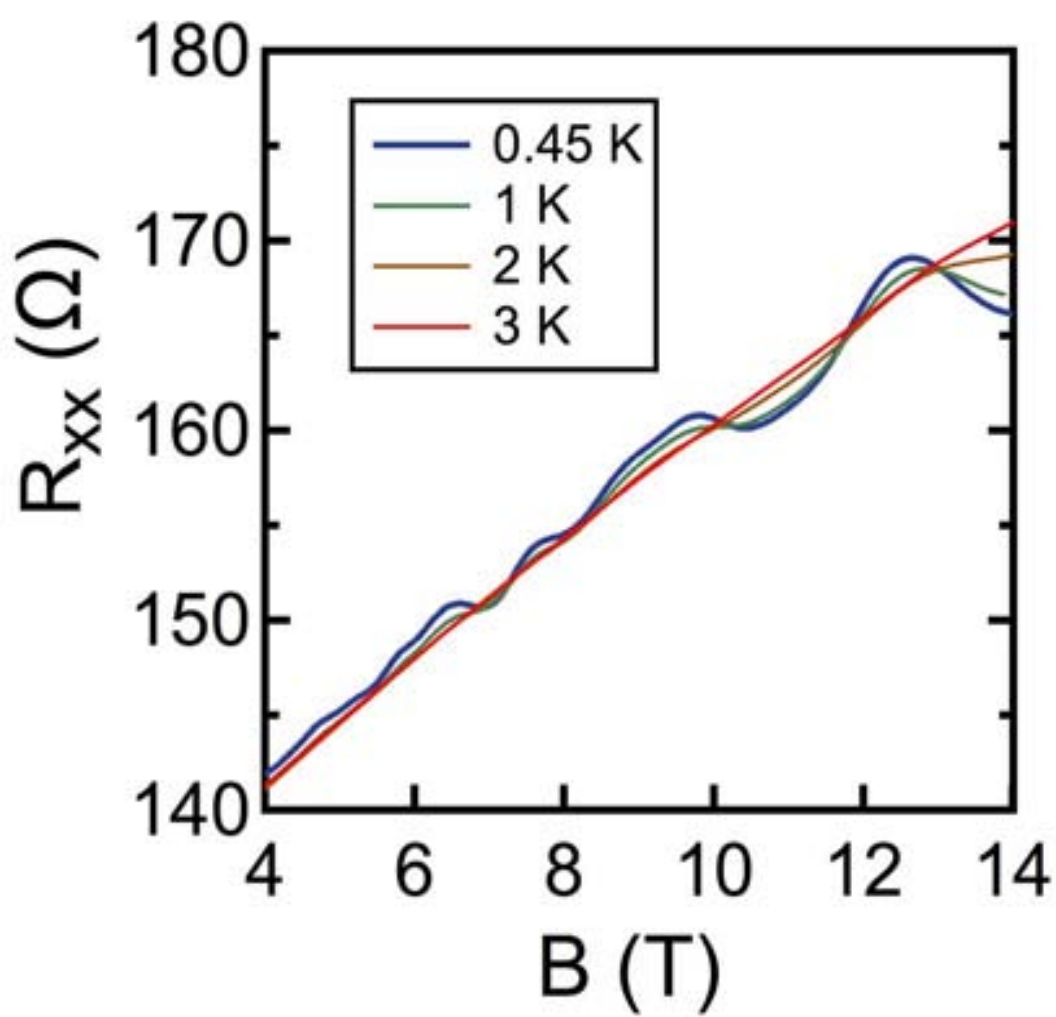

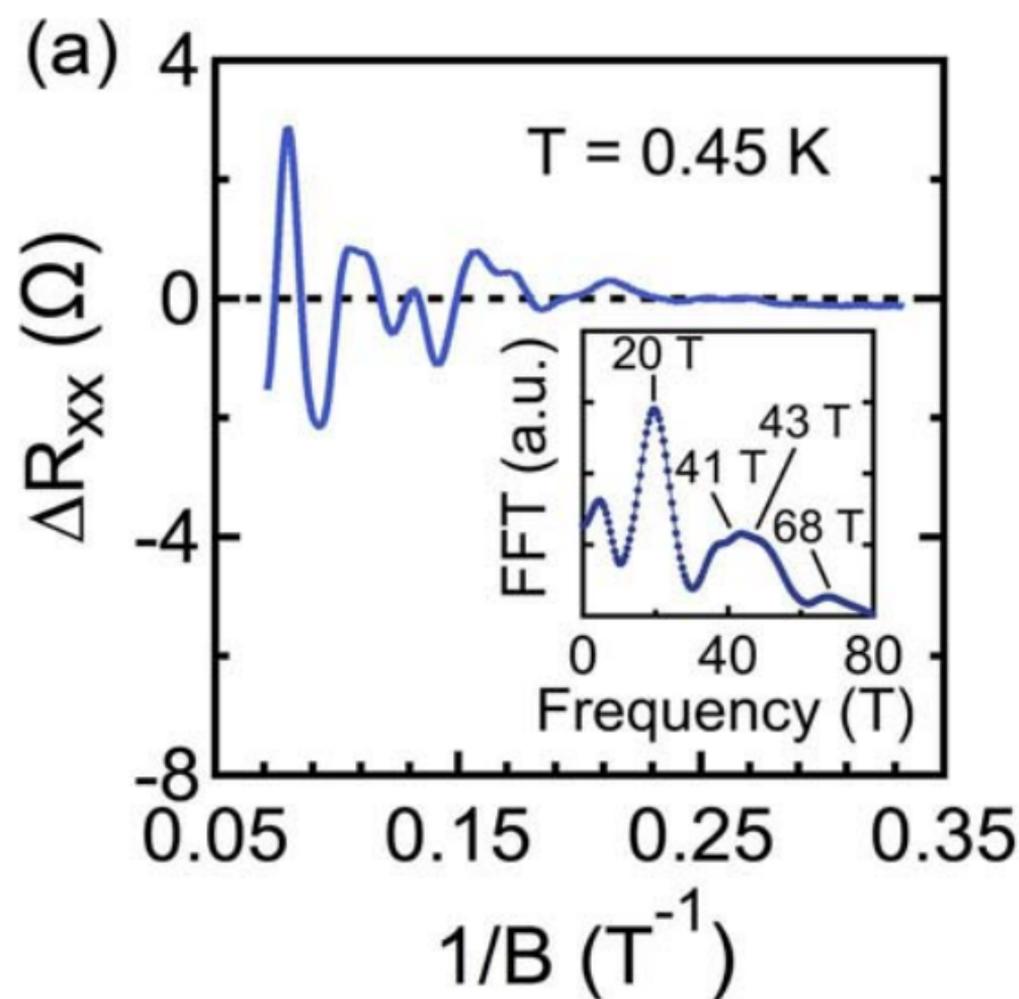
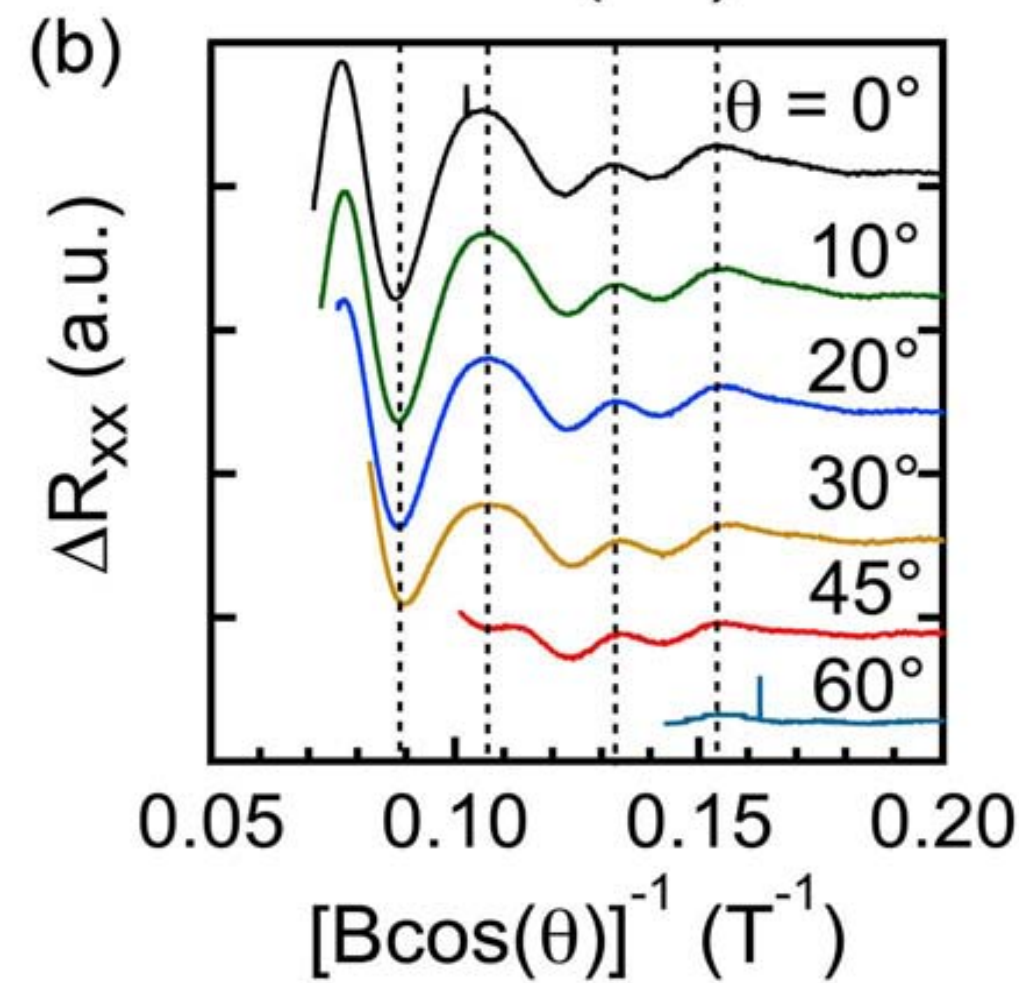

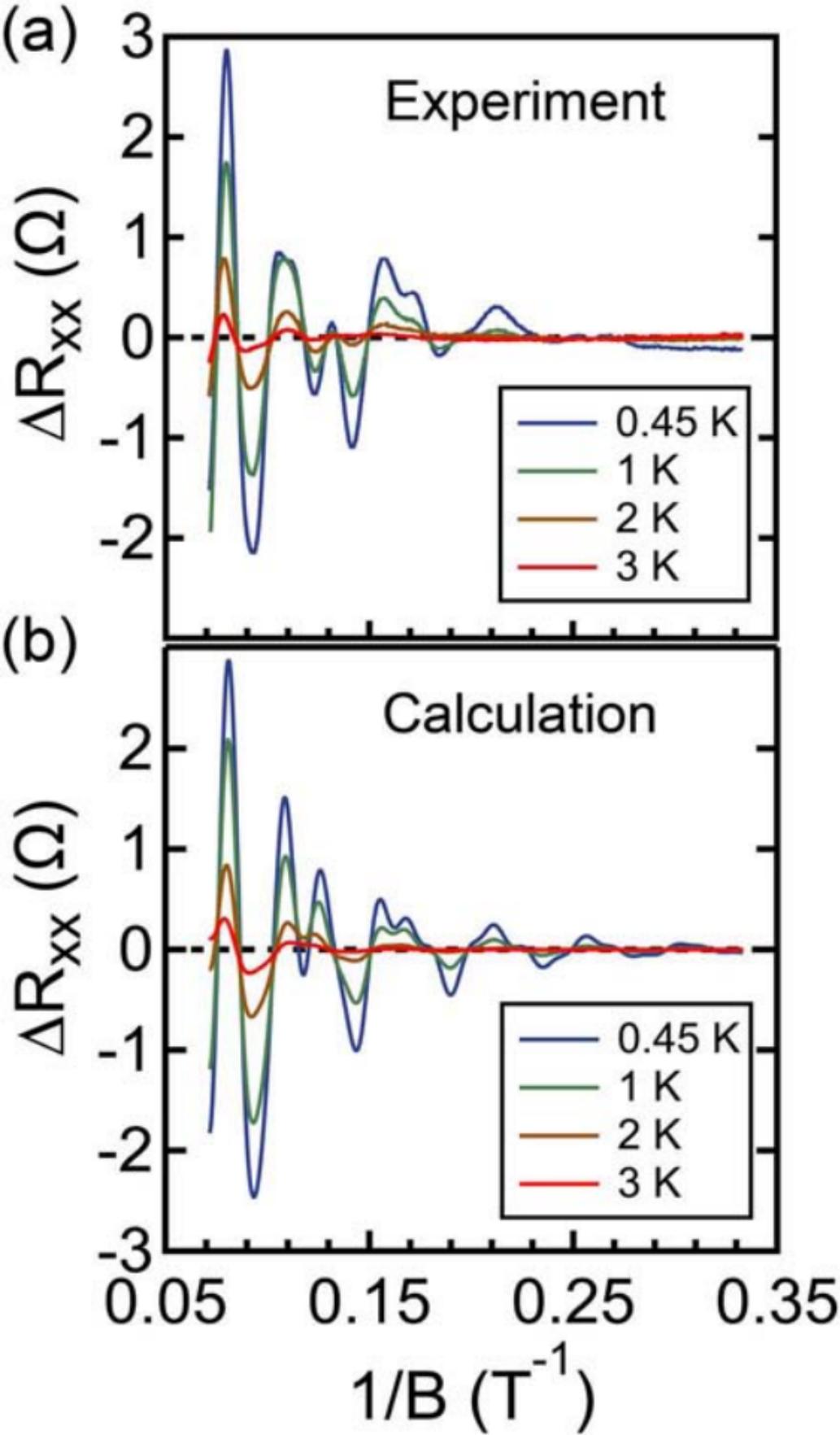

*Supplementary Information*

Two-dimensional electron gas in a modulation-doped SrTiO$_3$/Sr(Ti,Zr)O$_3$ heterostructure

Adam P. Kajdos[1], Daniel G. Ouellette[2], Tyler A. Cain[1], and Susanne Stemmer[1]

[1]Materials Department, University of California, Santa Barbara, California, 93106-5050, USA.
[2]Department of Physics, University of California, Santa Barbara, California, 93106-9530, USA.

**A. Growth and structural characterization**

SrTi$_{0.95}$Zr$_{0.05}$O$_3$ films were grown by hybrid molecular beam epitaxy (MBE), supplying both zirconium tert-butoxide (ZTB) and titanium tetraisopropoxide (TTIP) such that beam equivalent pressure (BEP) of ZTB comprised 12% of the sum of both metal-organic BEP's. All films were grown on (001) SrTiO$_3$ single crystal substrates. Growth was monitored using in-situ reflection high-energy electron diffraction (RHEED).

Figure S1 shows the intensity of specular 00 and first-order 10 RHEED streaks along the [100] azimuth as a function of time at the beginning of growth of a typical SrTi$_{0.95}$Zr$_{0.05}$O$_3$ film. Oscillations in the RHEED intensity indicate layer-by layer growth, and the decrease in the amplitude of the oscillations with time marks a transition to the step-flow growth mode, which is also observed in the growth of homoepitaxial SrTiO$_3$ by hybrid MBE [1].

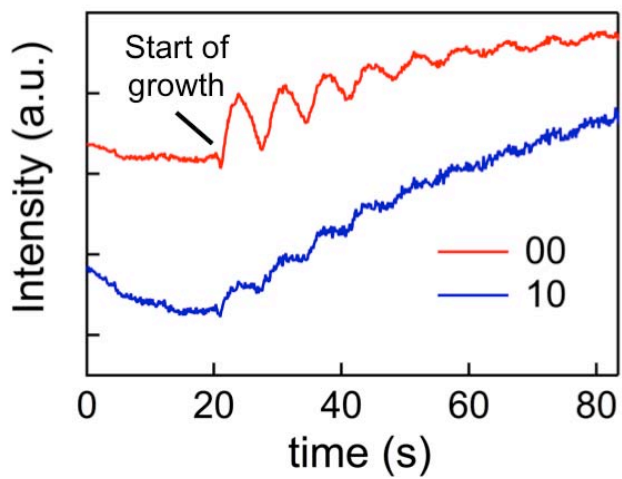

**Figure S1:** RHEED intensity as a function of time during growth of a typical SrTi$_{0.95}$Zr$_{0.05}$O$_3$ film.

Figure S2 shows RHEED patterns taken along the [100] and [110] azimuths of a thick (180 nm) La-doped ($n_{La}$ = 1.2 × $10^{20}$ $cm^{-3}$) $SrTi_{0.95}Zr_{0.05}O_3$ film grown on (001) $SrTiO_3$. Streaks in the RHEED pattern indicate that the film surface is atomically smooth. The first Laue zone is visible along the [110] azimuth, and ½-order reflections can be seen along the [100] azimuth, which is also observed in stoichiometric $SrTiO_3$ thin films grown by hybrid MBE [1].

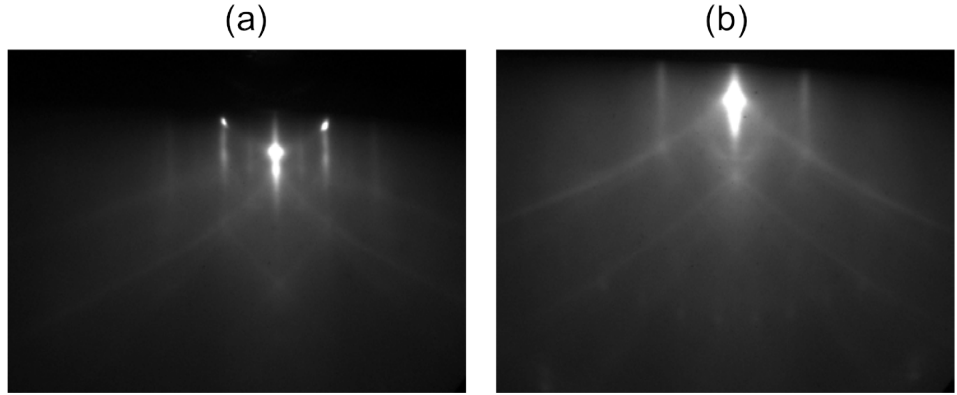

**Figure S2:** RHEED patterns taken along the (a) [100] and (b) [110] azimuths for a typical $SrTi_{0.95}Zr_{0.05}O_3$ film.

Figure S3 shows an on-axis high resolution x-ray diffraction (XRD) 2θ-ω scan of a 180 nm La:$SrTi_{0.95}Zr_{0.05}O_3$ film performed with a X'PERT Panalytical Pro Thin Film Materials Research Diffractometer equipped with a duMond–Hart–Partels Ge (440) monochromator. The $SrTi_{0.95}Zr_{0.05}O_3$ 002 Bragg peak is indexed in (pseudo)cubic unit cell notation. The room temperature crystal structure of the heteroepitaxial film is ambiguous at this composition [2]. The presence of a single $SrTi_{0.95}Zr_{0.05}O_3$ peak with clearly defined Laue oscillations indicates a solid solution film of high structural quality.

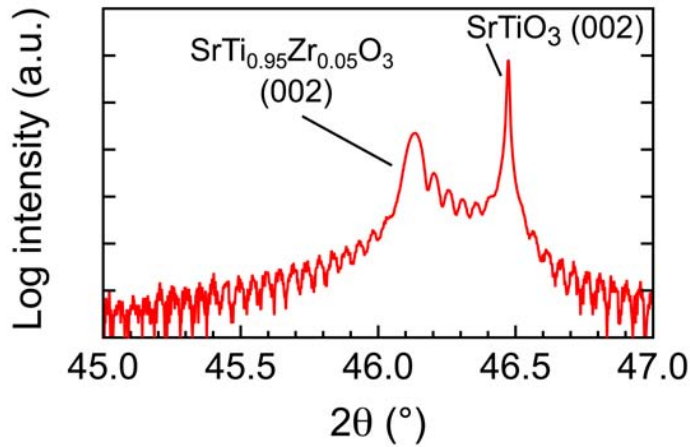

**Figure S3:** High-resolution XRD 2θ-ω scan of a 180 nm La-doped $SrTi_{0.95}Zr_{0.05}O_3$ film in the vicinity of the $SrTiO_3$ (002) substrate peak.

## B. Alternate fits to the experimental Shubnikov-de Haas oscillations

Other models for Shubnikov-de Haas oscillations were explored as fits to the experimental data in addition to the four-subband model discussed in the main text. Given that multiple peaks in the FTs are observed, one possibility is that they arise from only two subbands, but that the additional peaks are due to higher harmonics. Thus the following equation was used to fit the Shubnikov-de Haas oscillations:

$$\frac{\Delta R_{xx}}{R_0} = \sum_{i=1}^{l}\sum_{s=1}^{2} 2A_i \exp\left(-\frac{2s\pi^2 k_B T_{D,i}}{\hbar\omega_c}\right)\frac{sX}{\sinh(sX)}\cos\left(\frac{2\pi s f_i}{B}+s\pi\right), \qquad (S1)$$

where $i$ is the subband index, $l$ the number of occupied subbands ($l = 2$ in this case), $s$ is the order of the harmonic, $A_i$ are amplitude factors associated with intrasubband scattering probabilities, $k_B$ is the Boltzmann constant, $T_{D,i}$ is the Dingle temperature of each subband, $\hbar$ is the reduced Planck's constant, $\omega_c$ is the cyclotron frequency, $f_i$ are the oscillation frequencies (in $1/B$), and $X = \left(2\pi^2 k_B T\right)/\left(\hbar\omega_c\right)$. The parameters obtained from the fit are presented in Table SI, and the calculated $\Delta R_{xx}(1/B)$ is presented in Fig. S4.

**Table SI:** Parameters extracted from fits of the experimental Shubnikov-de Haas oscillations measured at 0.45 K to a two-subband model including first and second harmonic terms [Eqn. (S1), $l = 2$]. Also shown are the sheet carrier densities ($n_s$), quantum scattering times ($\tau_q$), and quantum mobilities ($\mu_q$) of each subband extracted from the fits. The effective mass $m^*$ is determined from the cyclotron frequency $\omega_c = eB/m^*$, $\tau_q$ is calculated from $\tau_q = \hbar/(2\pi k_B T_D)$, and $n_s$ is calculated from $f = n_s h/2q$, where $h$ is Planck's constant.

| Subband index $i$ | $R_0 A$ ($\Omega$) | $f$ (T) | $m^*$ ($m_0$) | $T_D$ (K) | $n_s$ (cm$^{-2}$) | $\tau_q$ (s) | $\mu_q$ (cm$^2$V$^{-1}$s$^{-1}$) |
|---|---|---|---|---|---|---|---|
| 1 | 10 | 42.2 | 1.08 | 1.5 | $2.04 \times 10^{12}$ | $8.10 \times 10^{-13}$ | 1320 |
| 2 | 2  | 21.4 | 1.02 | 0.8 | $1.03 \times 10^{12}$ | $1.52 \times 10^{-12}$ | 2620 |

Even though comparison of Fig. S4 and Fig. 4(a) indicates good qualitative agreement between experimental $\Delta R_{xx}(1/B)$ and fits using Eq. S1, the FT of the calculated data, shown in Figure S5, indicates that higher order harmonics are too weak to give rise to additional FT peaks that are clearly observed in the experiment [see Fig. 3(a)]. We therefore conclude that the experimental data are not due to only two subbands.

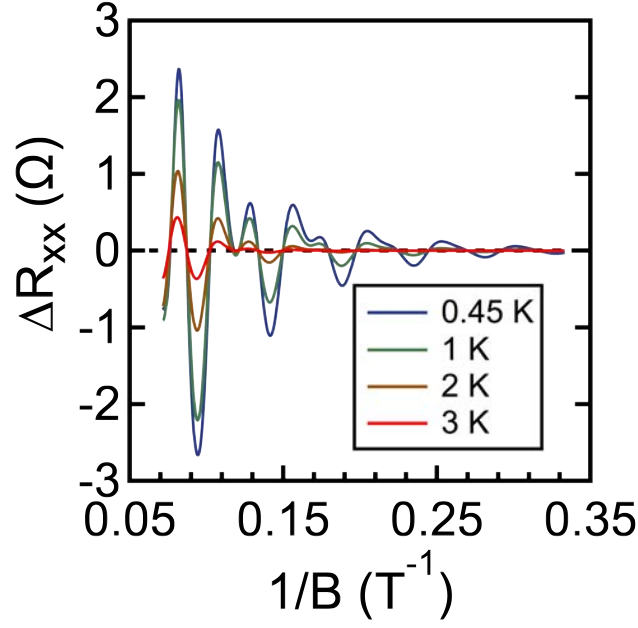

**Figure S4:** Shubnikov-de Haas oscillations calculated using Eq. S1, $l = 2$, and the parameters listed in Table SI.

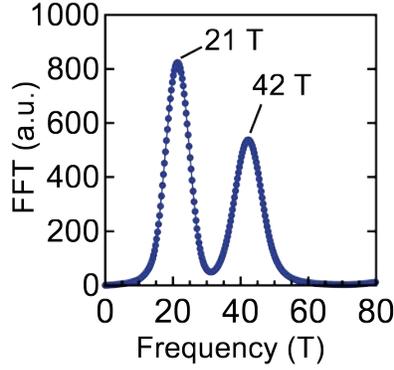

**Figure S5:** Fast Fourier transform (FFT) of calculated $\Delta R_{xx}(1/B)$ shown in Fig. S4. Only two frequencies are observed.

A second, alternative model involving two spin-split subbands was also explored. The data was fitted using the following equation:

$$\frac{\Delta R_{xx}}{R_0} = \sum_{i=1}^{l} 2A_i \exp\left(-\frac{2\pi^2 k_B T_{D,i}}{\hbar \omega_c}\right) \frac{X}{\sinh(X)} \cos\left(\frac{2\pi f_i}{B} + \pi\right) Z_s . \quad \text{(S2)}$$

The Zeeman term $Z_s$ is given as:

$$Z_s = \cos\left(\frac{\pi}{4}\frac{m^*}{m_e}g^*\right), \qquad (S3)$$

where $m^*$ is the effective mass, $m_e$ is the electron rest mass, and $g^*$ is the Landé factor. It is assumed that the magnetic field is oriented parallel to the sample surface normal, which is consistent with experimental procedure. The parameters obtained from the fit and the calculated $\Delta R_{xx}(1/B)$ are presented in Table SII and Figure S6, respectively. The calculated $\Delta R_{xx}(1/B)$ qualitatively agrees with the experimental data. The effective masses are lower than those expected for $d_{xy}$-derived subbands in SrTiO$_3$.

**Table SII:** Parameters extracted from fits of the experimental Shubnikov-de Haas oscillations measured at 0.45 K to a two-spin-split-subband model [Eq. S2, $l = 2$]. $n_s$ is calculated from $f = n_s h/q$, where $h$ is Planck's constant. Also see caption to Table SI.

| Subband index $i$ | $R_0 A$ (Ω) | $f$ (T) | $m^*$ ($m_0$) | $T_D$ (K) | $g^*$ | $n_s$ (cm$^{-2}$) | $\tau_q$ (s) | $\mu_q$ (cm$^2$V$^{-1}$s$^{-1}$) |
|---|---|---|---|---|---|---|---|---|
| 1 | 5 | 34.3 | 0.73 | 1.35 | 2.6 | $8.28 \times 10^{11}$ | $9.00 \times 10^{-13}$ | 2170 |
| 2 | 7 | 21.3 | 0.67 | 1.1 | 2.65 | $5.14 \times 10^{11}$ | $1.10 \times 10^{-12}$ | 2880 |

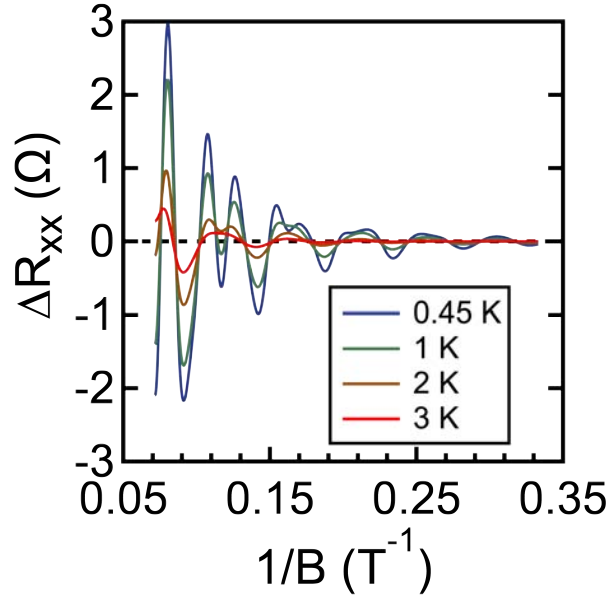

**Figure S6:** Shubnikov-de Haas oscillations calculated using Eq. S2, $l = 2$, with the parameters listed in Table SII.

## C. Discussion of the Hall coefficient

The sheet carrier density $n_{s,i}$ in a subband $i$ giving rise to a Shubnikov-de Haas oscillation with frequency $f_i$ can be determined from $f_i = n_{s,i} h/2q$, where $q$ is the electron charge and $h$ is Planck's constant. Using the values in Table I in the main text, the total sheet carrier density contributing to the Shubnikov-de Haas oscillations at 0.45 K ($n_{s,2DEG}$) from the four main $f_i$ [see inset in Fig. 3(a)] is approximately $8.4\times10^{12}$ cm$^{-2}$. The Hall carrier density at 300 K ($n_{s,H}$) is $4.46\times10^{13}$ cm$^{-2}$ [given by $n_{s,H} = 1/(q \cdot R_H)$, where $R_H$ is the Hall coefficient]. At room temperature, $n_{s,H}$ represents the true density, since all layers and subbands have a similar, low mobility limited by optical phonon scattering. The difference between $n_{s,2DEG}$ and $n_{s,H}$ can be explained with residual carriers in the La-doped SrTi$_{0.95}$Zr$_{0.05}$O$_3$ (also see band diagram in Fig. 1). A two-carrier interpretation of the zero-field Hall coefficient at 2 K supports this assertion, if we assume the following reasonable parameters for this sample. The carriers in the 2DEG have $n_{s,2DEG} = 8.4\times10^{12}$ cm$^{-2}$ and a mobility $\mu_{2DEG} = 2000$ cm$^2$V$^{-1}$s$^{-1}$ (i.e. the approximate average mobility in the subbands determined from the quantum scattering time, see Table I; estimating the transport mobility from the quantum mobility is reasonable when interface roughness scattering limits the mobility [3]). The carriers in the SrTi$_{0.95}$Zr$_{0.05}$O$_3$ then have a density of $n_{s,H} - n_{s,2DEG} = n_{s,STZO} = 3.66 \times 10^{13}$ cm$^{-2}$ and a mobility $\mu_{STZO}$ of 65 cm$^2$V$^{-1}$s$^{-1}$ ($\mu_{STZO}$ is consistent with values measured from a 180 nm La:SrTi$_{0.95}$Zr$_{0.05}$O$_3$ films isolated from the SrTiO$_3$ substrate by a 30 nm undoped SrTi$_{0.95}$Zr$_{0.05}$O$_3$ buffer). Using these values, the Hall coefficient can be estimated as a function of magnetic field $B$ [4]:

$$R_H = -\frac{1}{q}\frac{(\mu_{2DEG}^2 n_{s,2DEG} + \mu_{STZO}^2 n_{s,STZO}) + (\mu_{2DEG}\mu_{STZO}B)^2 (n_{s,2DEG} + n_{s,STZO})}{(\mu_{2DEG} n_{s,2DEG} + \mu_{STZO} n_{s,STZO})^2 + (\mu_{2DEG}\mu_{STZO}B)^2 (n_{s,2DEG} + n_{s,STZO})^2} \quad (S4)$$

The calculated value for the zero-field Hall coefficient $R_H(B=0) = 57.5$ m$^2$C$^{-1}$ compares favorably with experimentally determined value of $R_H(B=0) = 57.3$ m$^2$C$^{-1}$ at 2 K. The Hall resistance $R_{xy} = R_H B$ calculated from Eqn. (S4) remains linear to 14 T, which is in good agreement with the measured $R_{xy}$ (see Figure S7), which is only weakly nonlinear to 14 T at 2 K.

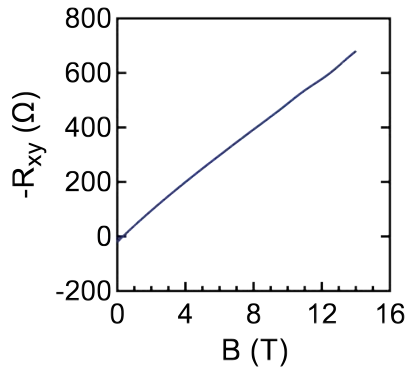

**Figure S7:** Hall resistance R$_{xy}$ of the modulation doped heterostructure measured at 2 K.